\documentclass[preprint,aip,jcp,amsmath,amssymb,floats,superscriptaddress]{revtex4-1}
\usepackage{graphicx}
\usepackage{color}

\begin{document}
\title{How Close to Two Dimensions Does a Lennard-Jones System Need to Be to Produce a Hexatic Phase?}
\author{Nadezhda Gribova}
\email{gribova@icp.uni-stuttgart.de}
\affiliation{Institute for Computational Physics, University of Stuttgart, Pfaffenwaldring 27, D-70569 Stuttgart, Germany} 
\affiliation{present address: Institute of Thermodynamics and Thermal Process Engineering, University of Stuttgart, Pfaffenwaldring 9, D-70569 Stuttgart, Germany}
\affiliation{Institute for High Pressure Physics, Russian Academy of Sciences, Troitsk 142190,
Moscow Region, Russia}
\author{Axel Arnold}
\affiliation{Institute for Computational Physics, University of Stuttgart,Pfaffenwaldring 27, D-70569 Stuttgart, Germany}
\affiliation{previous address: Fraunhofer SCAI, Schloss Birlinghoven, D-53754 Sankt Augustin, Germany}
\author{Tanja Schilling}
\affiliation{Universit{\'e} du Luxembourg, 162 A, avenue de la Fa\"{\i}encerie, L-1511 Luxembourg }
\author{Christian Holm}
\affiliation{Institute for Computational Physics, University of Stuttgart,Pfaffenwaldring 27, D-70569 Stuttgart, Germany}
\date{\today}

\begin{abstract}
We report on a computer simulation study of a Lennard-Jones liquid confined in a narrow slit pore with 
tunable attractive walls. In order to investigate how freezing in this system occurs, we perform 
an analysis using different order parameters. Although some of the parameters 
indicate that the system goes through a hexatic phase, other parameters do not. This
shows that to be certain whether a system has a hexatic phase, one  needs to study not only a large system, 
but also several order parameters to check all necessary properties. 
We find that the Binder cumulant is the most reliable one  to prove the existence of a hexatic phase.
We observe an intermediate hexatic
phase only in a monolayer of particles confined such that the fluctuations in the positions
  perpendicular to the walls are less then $0.15$ particle diameters, i. e. if the system is practically perfectly 2d.
\end{abstract}

\pacs{}
\maketitle

\section{Introduction}
Understanding the structure and dynamics of confined fluids is
important for processes such as wetting, coating, and nucleation. The
properties of a fluid confined in a pore differ significantly from the
bulk fluid due to finite size effects, surface forces and reduced
dimensionality. In this work we report on a study of one of the simplest
models that is still capable of reproducing the thermodynamic behavior
of classical fluids, the Lennard-Jones (LJ) system. The LJ potential
is an important model for exploring the behavior of simple fluids and
has been used to study homogeneous vapor-liquid, liquid-liquid and
liquid-solid equilibrium, melting and freezing \cite{Alba2006,gasser09a,gasser10a}. It has also been used
as a reference fluid for complex systems like colloidal and polymeric
systems.

The vapor-to-liquid transition in confined systems has been studied
intensively, and is well understood (see~\cite{Gelb_review1999} and
references therein).  In this article we will discuss the
liquid-to-solid transition in a slit pore and the process of the
development of the solid phase.  In the liquid phase, confinement to a
slit induces layering at the walls.  One could imagine this effect to
facilitate crystallization. And indeed it is known that depending on
the strength of the particle-wall interaction, the freezing scenario
changes significantly \cite{Miyahara1997,Alba2006}.  If the walls are
strongly attractive, crystallization starts from the walls and at a
temperature higher than without confinement. If the walls are strongly
repulsive, crystallization starts from the bulk at a temperature lower
than without confinement.

A well-distinguished layer of particles close to the wall can also, to
some extent, be treated as a 2d system. The melting of true 2d systems has been studied both
theoretically~\cite{ryzhov95,ryzhov2002,Mak2006,lee08a,binder02a} and
experimentally~\cite{murray87,Marcus96,han08a}. 
A large number of experiments on 2d melting were carried out in colloidal
systems where colloidal particles contain a magnetic
core, giving rise to a magnetic repulsion between particles that can be
controlled by an external magnetic field (see, for example, 
\cite{zahn1998,zahn00a,keim07a}).
The type of the scenario strongly depends on the shape of the potential.  Soft-core
potentials melt via the Kosterlitz-Thouless-Halperin-Nelson-Young
(KTHNY) mechanism~\cite{ryzhov2002,lee08a}, meaning that the liquid
turns into a crystal going through an intermediate hexatic phase
\cite{KT73,halperin78,nelson79,young79,strandburg88a}. For the hard
disks system two different points of view
exist~\cite{ryzhov2002,binder02a,Mak2006,rice09a}.  Since the Lennard-Jones potential
is rather soft, the freezing of a single layer of LJ particles can
therefore be expected to proceed via the KTHNY mechanism \cite{wierschem10a}, which
significantly differs from the bulk nucleation scenario. 

As we will show, it can be difficult to check whether the hexatic
phase exists. To solve this problem several order parameters 
to characterize the bond-order were introduced in the literature. 
The correlation function of the local
bond-order parameter \cite{strandburg88a} that measures the nearest-neighbor-bond-angular
order is commonly used. However, it can not distinguish between a hexatic phase and 
a heterogeneous system in the two-phase region.
The distribution of the bond-angular susceptibility on various length scales was
introduced to overcome this problem studying a hard disks system and 
Lennard Jones disks \cite{strandburg84a}. Later the search for a general and
efficient method to determine  all phases and bounds of the transition was 
continued. The scale analysis of the behavior of the fluctuation of the  bond-angular 
susceptibility and the bond-orientation cumulant provided \cite{weber95a} an evidence of a possible 
continuous transition in the system of hard disks \cite{binder02a}. 
To our knowledge, the Binder cumulant was applied in the analysis  of the
existence of the hexatic phase only for 2d systems and never for quasi-2d or 3d.
The analysis of fluctuations of the  bond-angular susceptibility within a layer was used already for
studying the melting of thin films up to 20 layers \cite{peng10a}.
A modified scaling analysis of the bond-angular susceptibility \cite{bagchi96a}
was used not to check the existence of the hexatic phase only in 2d systems 
\cite{bagchi96a,jaster04a,Mak2006}, but also in quasi-2d systems,
for example \cite{Radhakrishnan2002}.

 This raises the question, how the crystallization
 in a strongly confined quasi 2d system proceeds, i.~e. a system only
 a couple of particle diameters wide. In this case, it is not clear
 whether the system still behaves like being truly two dimensional,
 or whether it rather behaves similar to a bulk system.
 The solid-solid phase transitions of confined fluids in narrow slit pores 
were studied both at zero  and at finite temperatures 
(see References \cite{bock2005,ayappa2007,kahn09} and references therein).
For a confined LJ fluid, the question if a hexatic phase exists in 
a quasi 2d system has been studied by Radhakristan
and coworkers \cite{radhakrishnan2002JCP} for a ratio of wall-particle
to particle-particle attraction varying between 0 and 2.14 and pores widths
of 3 and 7.5 fluid particle diameters. For the narrower slit
pore it was shown that around the freezing temperature the system
exhibits a hexatic phase.  With increasing wall attraction this
temperature region becomes wider, i.~e.~an attractive wall facilitates
the formation of the hexatic phase. The phase diagram for the wider
pore with diameter 7.5 is more complicated. When the wall-particle
attraction becomes bigger than the particle-particle attraction, at
first a hexatic phase and then a crystal phase appear, however, only
in the contact layers near the walls; the rest of the system remains
liquid. Only when decreasing the temperature further the system
crystallizes completely. The temperature range, in which hexatic or
crystal phases are observed only in the contact layers, again widens
with growing wall-particle attraction. This indicates that the
wall-particle attraction facilitates the formation of a hexatic phase
even in wider pores, however only in the layers close to the walls.
The same group of authors also reported that in a pore, that can
accommodate only a single layer, two second order transitions are observed, 
while already in a pore wide enough to accommodate two layers,
both transitions are of the first order ~\cite{Radhakrishnan2002}.

The different crystal structures of the frozen phase were studied by Vishnyakov and Neimark
\cite{vishnyakov2003} as a function of the size of the slit for this
system. The distance between the walls was gradually increased up to a
slit accommodating three layers. Depending on the width of the pore,
hexagonal or orthorhombic phases were observed in the layers.

In a recent article by Page and Sear~\cite{page2009} it was shown that 
freezing is controlled  by prefreezing in a similar system.
Nucleation of the bulk crystal is affected by the surface phase
behavior. With increasing wall attraction, the bulk nucleation is
smoothly transformed into nucleation of a surface crystal layer. 
Xu and Rice\cite{xu2008} investigated theoretically a
quasi 2d system of hard spheres and reported the dependence of the
density at the liquid-to-hexatic phase transition on the thickness
of the system, with wall separation changing from 1 to 1.6 hard
sphere diameters.
For the current state of art in crystallization of confined
  systems we recommend to consult recent reviews
  \cite{gasser10a,gasser09a,rice09a}.

In this paper we study the influence of the confinement on the hexatic phase.
We investigate a Lennard-Jones fluid at different values of wall-particle
attraction during freezing and melting. The system is confined in an
attractive slit pore with changing wall separation, being able to
accommodate 1 or 2 layers.
To characterize and distinguish
the liquid, hexatic and solid phases we investigated several order parameters and
compared their behavior. We show that the identification of a hexatic phase is
depending on the order parameters one uses. 
There is some controversy in experiments regarding the observation of a
hexatic phase~\cite{marcus97a,karnchanaphanurach00a}. We study how strong the system
has to be confined to observe a hexatic phase, and whether such a phase can be 
observed also in multilayer systems, as predicted by Radhakrishnan et al~\cite{Radhakrishnan2002,radhakrishnan2002JCP}.

The article is structured as follows. In section \ref{sec:sim} we
describe our simulation method, and in section \ref{sec:results} we
present the order parameters and the results. 
We conclude with a summary \ref{sec:conclusions}.

\section{Simulation method}\label{sec:sim}

We performed molecular dynamics (MD) simulations of Lennard-Jones 
particles confined between two structureless walls. The particles 
interact via the LJ-potential
\begin{equation}
u(r)=4 \epsilon \left[ \left( \frac{\sigma}{r}\right)^{12}-\left(\frac{\sigma}{r}\right)^{6}\right],
\end{equation}
where $r$ is the distance between the particles, $\sigma$ the
particle diameter and $\epsilon$ the depth of the minimum of the LJ
potential. The interaction between walls and particles is given by a
LJ-potential integrated over semi-space:
\begin{equation}
 u_w(r)=4\epsilon_w\left[ \left( \frac{\sigma}{r}\right)^{9}-\left(\frac{\sigma}{r}\right)^{3}\right].
\end{equation}
The particle-particle interaction was cut off and shifted at a distance $r_c=2.5
\sigma$ and the wall-particle interaction at a distance
$r_c=4.0\sigma$, since the wall-particle potential is wider and deeper
than the particle-particle potential. For the following we will use $\epsilon$ as the
unit of energy, $\sigma$ as the unit of length and $\tau=\sqrt{1 \cdot
  \sigma^{2}/\epsilon}$ as unit of time (i.~e.~use the particle mass
as the unit of mass); consequently, temperatures are given in
multiples of $\epsilon/k_{\mathrm{B}} T$.  The simulations were
performed in a cuboid box with periodic boundary conditions in the $x$-
and $y$- directions and two walls positioned at $z=0$ and $z=L_z$. The
distance between the walls was chosen such that $n=1,2$ layers can
be accommodated in the pore, namely $L_z=2 \cdot 1.12+0.916\cdot
(n-1)$. Here, 1.12 is the distance at which the wall potential has its
minimum, and 0.916 is the layer distance in an ideal FCC lattice with
spacing one. Therefore $L_z$ was either 2.24 or 3.16 for one or two
layers respectively, while the
other two dimensions of the simulation box were fixed as $L_x=L_y=200$.
The number of particles $N$ was chosen such that the density was kept
constant at one particle per unit cube independent of the width of the
slit, and therefore ranged from  44800 for one layer to 81600 particles
for two layers. Since the slit is narrow, layering in the two 
layered system is observed in the whole range of the temperatures.

We carried our simulation out in the NVT ensemble, since this corresponds
to the way recent experiments were done \cite{han08a, peng10a}, although 
our parameters do not strictly allow to reproduce these experiments.
The simulations ran $1.0\times10^{6}$ MD steps for equilibration and
$2.5\times10^{5}$ MD steps for sampling.  
For our simulations the software package ESPResSo version
2.1.2j was used \cite{limbach06a}.

\section{Results and Discussion}\label{sec:results}

To check whether the system is in a hexatic phase one usually studies  the
decay of the  $G_6$ correlation of the local bond-order parameter,
the orientational susceptibility $\chi_6$: the scaling of its mean for different system sizes \cite{bagchi96a}
and its probability distribution for different system sizes \cite{strandburg84a}, to
check the homogeneity of the system and for finite size effects.
In our work we also study the scaling of a modified susceptibility $\chi'_6$ (fluctuation) for different system sizes 
and temperatures \cite{lee92a,weber95a,han08a} and (to our knowledge used only in strictly 2d systems before) the
Binder cumulant of $\psi_6$ \cite{binder81a,weber95a}.

We would like to introduce these parameters at the example of the system accommodating one layer and  
with the wall attraction $\epsilon_w=5$.
All parameters are based on the local bond-order correlation parameter
since the hexatic phase is characterized by quasi long-range bond order. 
The local bond-order correlation parameter 
of particle $j$ in layer $m$ at a position $\mathbf{x}_{j}$ 
is defined as
\begin{equation}
 \psi_{6}(\mathbf{x}_{j})=\frac{1}{N_{j}}\sum_{k=1}^{N_{j}}e^{i6\theta_{jk}}
\end{equation}
where $N_{j}$ is the number of neighbors of particle $j$ within layer m, 
the sum is over the neighbors $k$ of $j$ within $m$, 
and $\theta_\textit{jk}$ is the angle between
an  arbitrary fixed axis and the line connecting particles j and k. 
The order parameter of the layer $\Psi_6$ is defined as the average over
$\psi_{6}(\mathbf{x}_{j})$ for all $N$ particles within the layer
\begin{equation}
 \Psi_6=\frac{1}{N}\left|\sum^{N}_{j=1}\psi_{6}(\mathbf{x}_{j})\right|.
\end{equation}

The correlation function $G_6(r)$ of the local bond-orientational order
helps to distinguish long- and short-range orientational order.
It is defined as
\begin{equation}
G_{6}(r)=\langle\psi^{*}_{6}(\mathbf{x}^{'})\psi_{6}(\mathbf{x})\rangle
\quad ,
\end{equation}
where the average is taken over all particles within a layer where
positions $\mathbf{x'}$ and $\mathbf{x}$ are a distance $r$ apart.

\begin{figure}[t]
\includegraphics[width=8 cm]{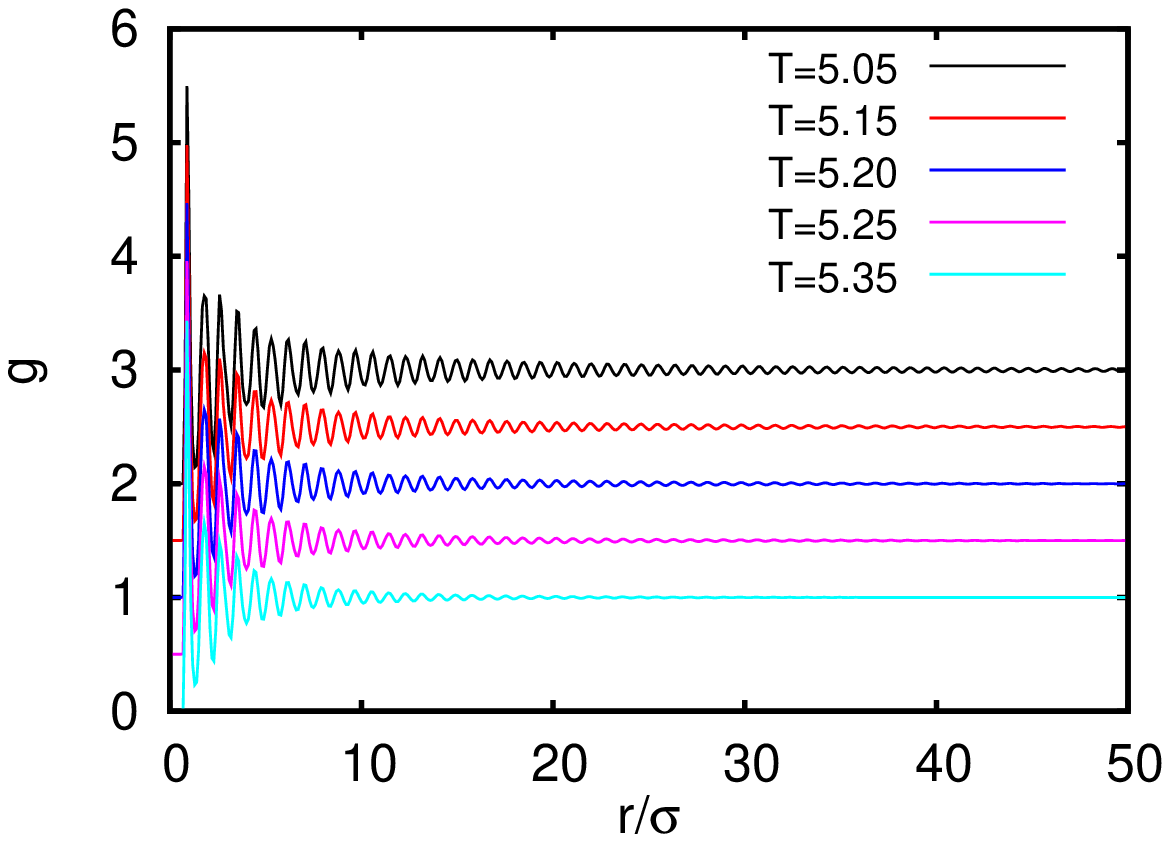}
\includegraphics[width=8 cm]{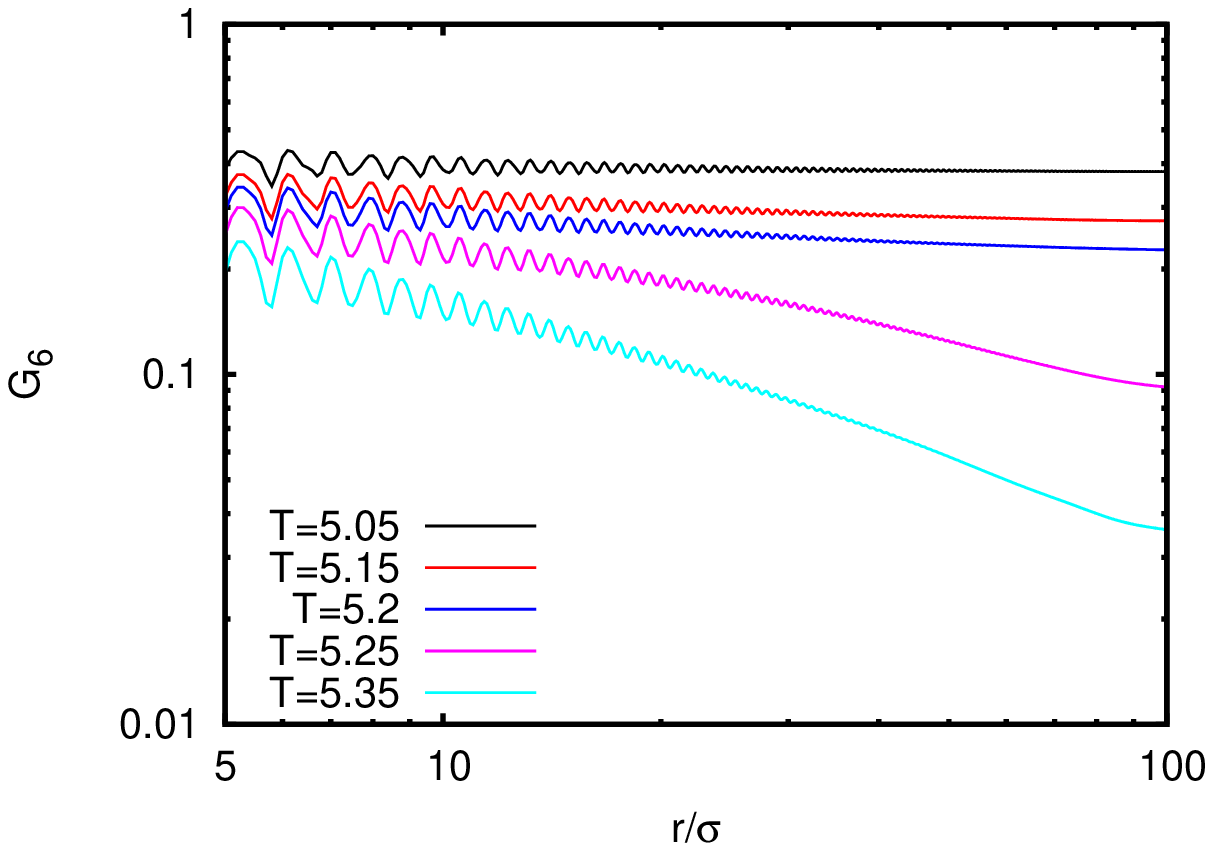}
\caption{Left figure: Radial distribution function $g(r)$ for one layer,
  $\epsilon_w=5$ versus $r$. The curves are
shifted along the y-axis to separate them. The RDF for $T=5.05, 5.15$  is quasi long-ranged 
and solid-like, and for  $T=5.2, 5.25$ and $5.35$ it is short-ranged as for liquid.\\
Right figure: Correlation function $G_6(r)$ of the bond order parameter versus
$r$. $G_6$ does not decay for $T=5.05$, 
for $T=5.15$ and $5.2$ it decays algebraically, for $T=5.25$ and  $5.35$ $G_6$
decays exponentially.} 
\label{fig:bond}
\end {figure}

The radial
distribution function in turn allows to distinguish long- and
short-range translational order and it is defined as
\begin{equation}
g(r)=\langle\rho(r)\rangle/\rho\quad,
\end{equation}
where $\langle\rho(r)\rangle$ denotes the average local density at
distance $r$ from a fixed particle, and $\rho$ the overall average
density.

Ideally, the decay of $G_6$ together with the radial distribution function $g(r)$ (RDF) 
allow to detect a hexatic phase.
 For a two-dimensional crystal with
long-range orientational and translational order, $G_6$ flattens to a
nonzero constant, while $g(r)$ decays very slowly to 1. In a two
dimensional liquid we have only short-range order, and therefore both
functions decay exponentially to 0 and 1, respectively. In the hexatic
phase with its long-range orientational, but short-range translational
order, $G_6$ decays algebraically, while $g(r)$ decays exponentially.

As we can see in the Fig.\ref{fig:bond}, the radial distribution for $T=5.05, 5.15$  is still quasi long-ranged 
and solid-like, and for  $T=5.2, 5.25$ and $5.35$ the RDF looks like the one for liquid. Meanwhile, $G_6$
only for $T=5.05$ does not show any decay, for $T=5.15$ and $5.2$ it decays algebraically and starting with $T=5.25$ 
it decays exponentially. Combining conclusions from
RDF and $G_6$ one can suspect that around $T=5.2$ there is a hexatic phase and at $T=5,15$ we have
a defective crystal.  We would also like
to note that the crossover to exponential decay at $T=5.25$ happens at long distances above 20, that
means that to make an appropriate judgment about type of decay, one needs sufficiently large boxes.

However, both $G_6$ and the RDF are averaged over the whole system, so if the system is not homogenous,
they can not detect that.
The obvious way to check the homogeneity, except for a direct observation, is
to divide the original system into several
subsystems and to compare the behavior of the parameters in each of subbox. 
Calculating $G_6$ in subsystems is not favorable,
since we are interested in the long range decay that becomes impossible to study with decreasing system size.

\begin{figure}[t]
\includegraphics[width=8 cm]{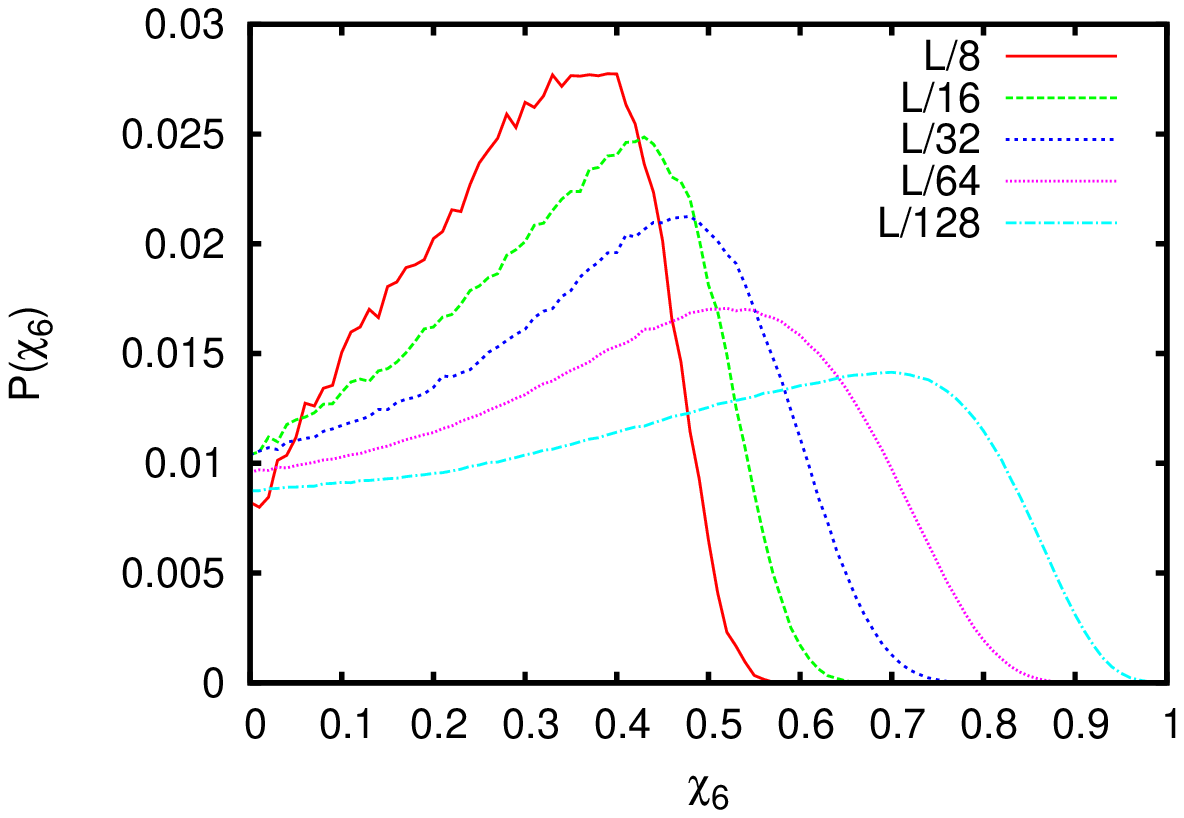}
\includegraphics[width=8 cm]{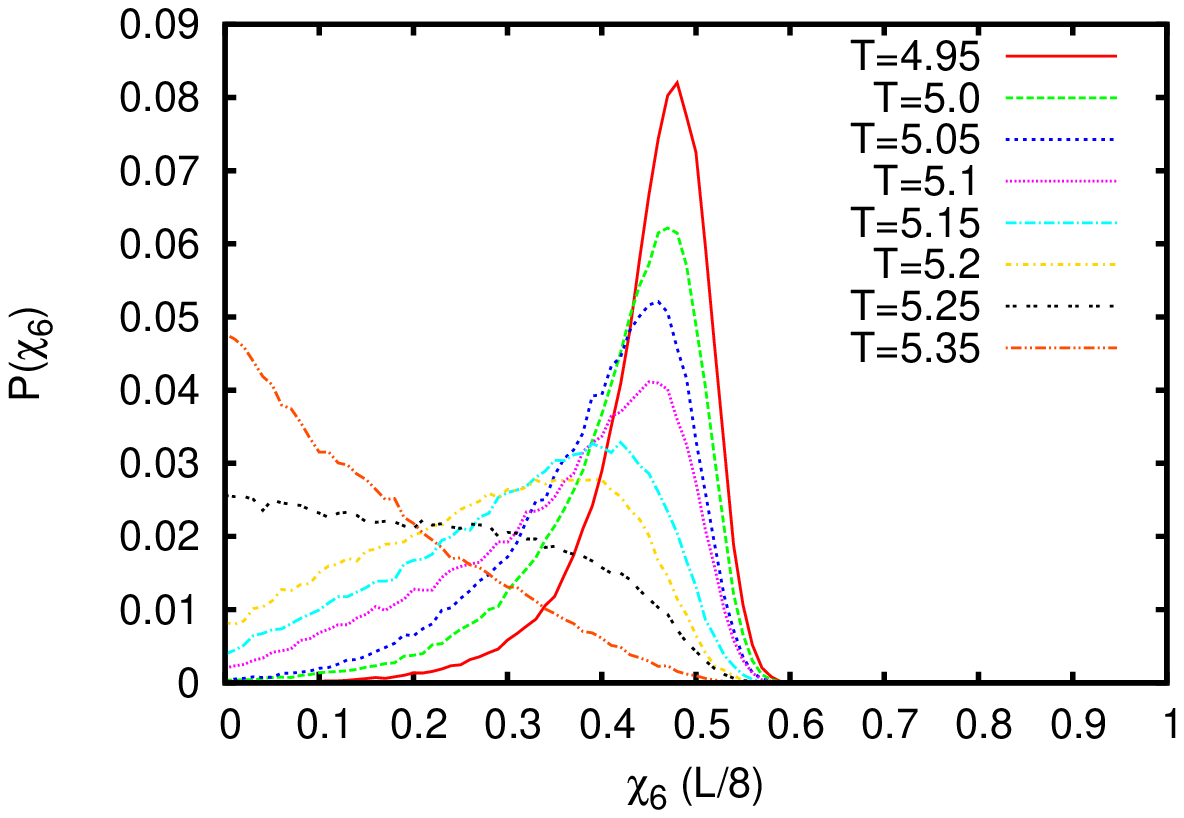}
\caption{Left figure: The distribution of $\chi_6$ for different subdivisions of the box at $T=5.2$ 
in one layer, $\epsilon_w=5$, where we suspected a hexatic phase. The distribution has one peak and the phase is homogeneous. \\
Right figure: The distribution of $\chi_6$ for subdivision $1/8$ at different temperatures. The peak of the
distribution gradually shifts away from $0.5$ and
becomes wider with increasing temperature, then at $T=5.25$ the system changes into the liquid state } 
\label{fig:distr1l}
\end {figure}

In reference \cite{strandburg84a} a study of the nearest-neighbor bond-angular susceptibility on various length scales 
was performed. The susceptibility is defined  as
\begin{equation}
 \chi_6=\left\langle \left|\frac{1}{N} \sum_j \psi_{6}(\mathbf{x}_{j}) \right|^2\right\rangle=
\langle \Psi^2_6\rangle
\label{chi1}
\end{equation}
This quantity is used to examine the system on different length scales by
dividing the system into equal subblocks of length
$L_b=L/2,L/4...L/64, L/128$, where L is the box length of the original
system. For each subblock the distribution of $\chi_6$
was computed. In the solid $\chi_6$ will be between 1 and 0.5 due 
to presence of the long range order and in the liquid 
$\chi_6$ will be close to zero. In the Fig.\ref{fig:distr1l} (left)  we show the distribution of $\chi_6$ at $T=5.2$, 
where hexatic phase is suspected. One can see that the distribution has one peak, so the phase should be homogeneous
at this temperature. The $\chi_6$ peak gradually shifts away from $0.5$ and
becomes wider with increasing temperature and then, at $T=5.25$, the system changes into the liquid state 
(Fig. \ref{fig:distr1l}(right)). Again, this supports the idea that at $T=5.2$ the system goes through a hexatic phase.

\begin{figure}[t]
\includegraphics[width=8 cm]{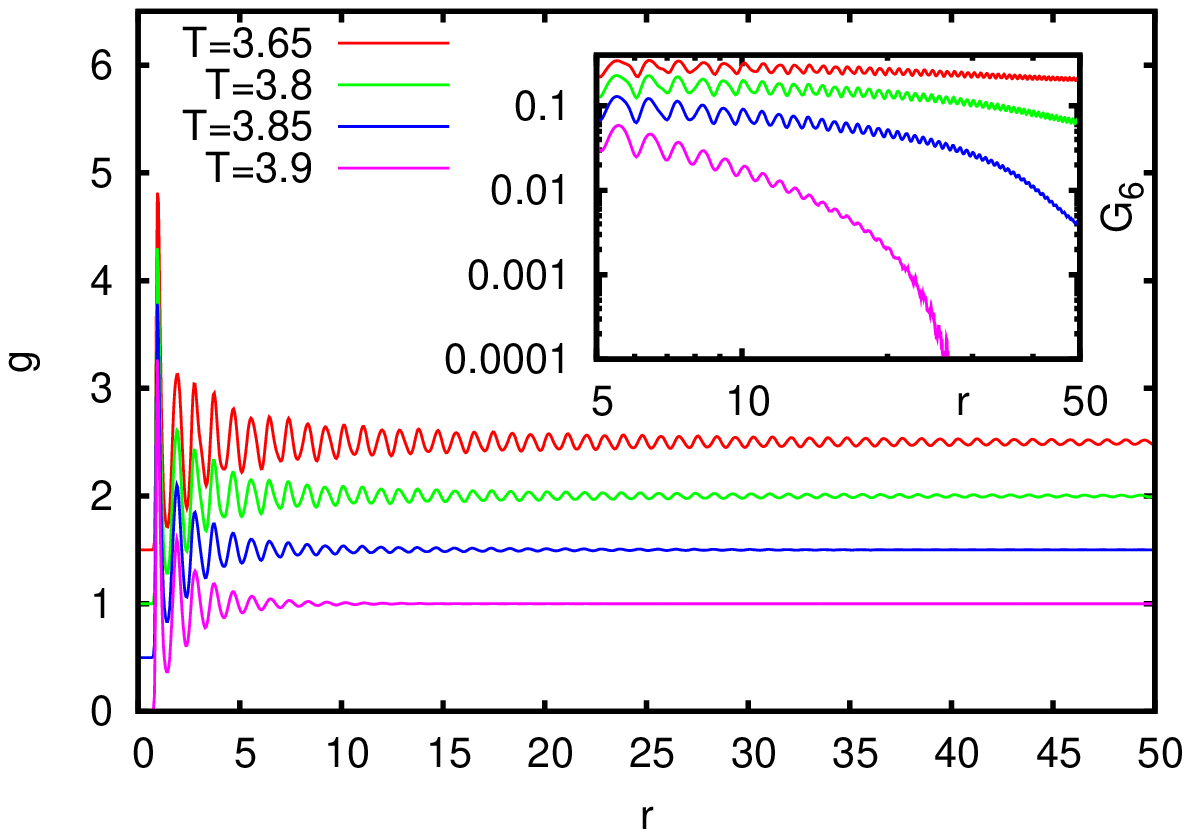}
\includegraphics[width=8 cm]{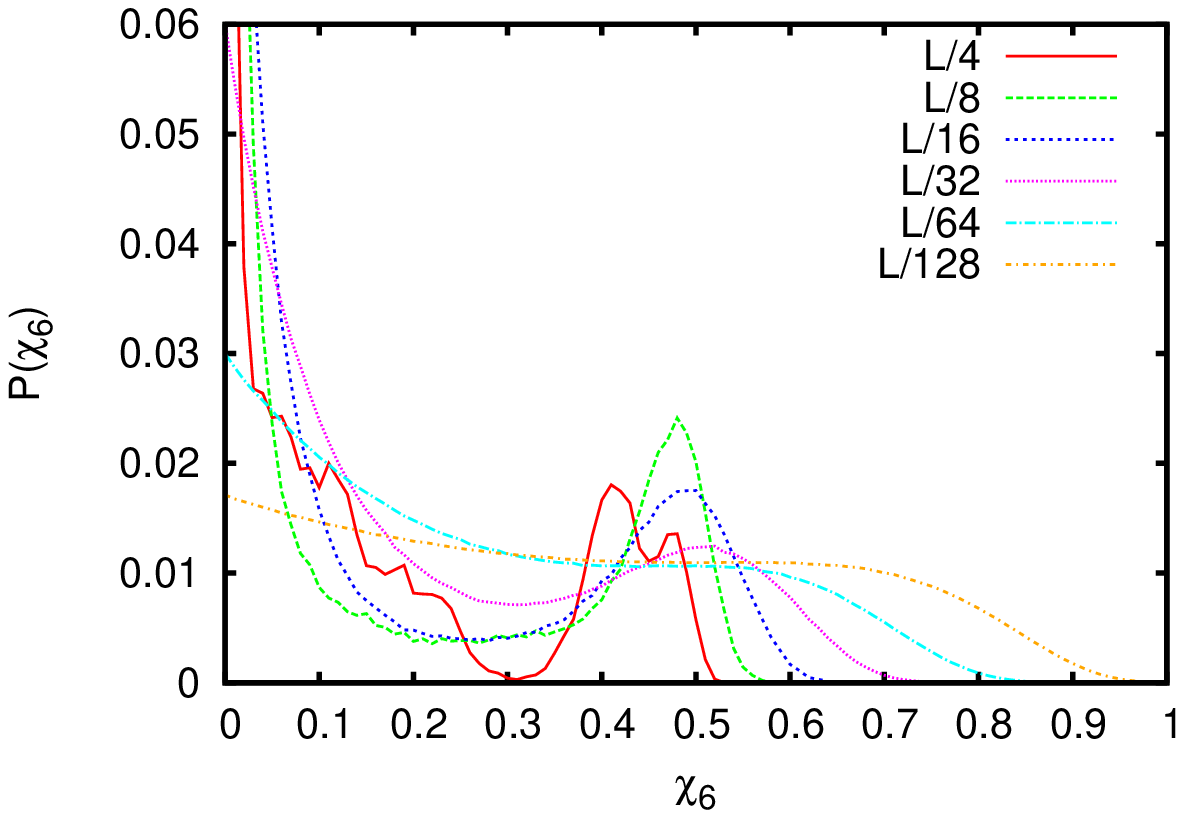}
\caption{Left figure: In 2 layers the radial distribution function for $T=3.65$ and $T=3.8$ still behaves solid-like and
for $T=3.85$ and $T=3.9$ it shows the liquid-like behavior.\\
Inset: Correlation  of the bond order parameter $G_6$  decays algebraically up to $T=3.8$, for $T=3.85$ 
it has at first an algebraic decay and that becomes later exponential, at  $3.9$ $G_6$
decays exponentially.\\
Right figure: The distribution of $\chi_6$ for different subdivisions of the box at $T=3.8$ 
in two layers, $\epsilon_w=5$. The distribution has two peaks around $0.5$ and $0$, that shows that 
the system has a liquid-solid coexistence.} 
\label{fig:distr2l}
\end {figure}

If the system is inhomogeneous, a combination of solid and liquid distribution
for subblocks with small length is observed, a vivid example is shown in Fig.\ref{fig:distr2l}(right) 
for the two layers system at $T=3.8$. In all two layer systems for different wall attraction we could see only 
three regimes: solid, liquid or  a solid-liquid coexistence. As expected, the
radial distribution function and the correlation of the bond-order parameter do not capture this (Fig.\ref{fig:distr2l}(left)). 
The RDF of up to $T=3.8$ is still quasi-long-range and then at $T=3.85$ it becomes short-range. 
The decay of the $G_6$ at $T=3.8$ is algebraic, at $T=3.85$ it shows a combination of algebraic and exponential decays and at $T=3.9$ it
becomes purely exponential (see the insert Fig.\ref{fig:distr2l}(left)). No homogeneous phase in the intermediate region was observed
for any of the wall-particles attractions we studied.

\begin{figure}[t]
\includegraphics[width=8 cm]{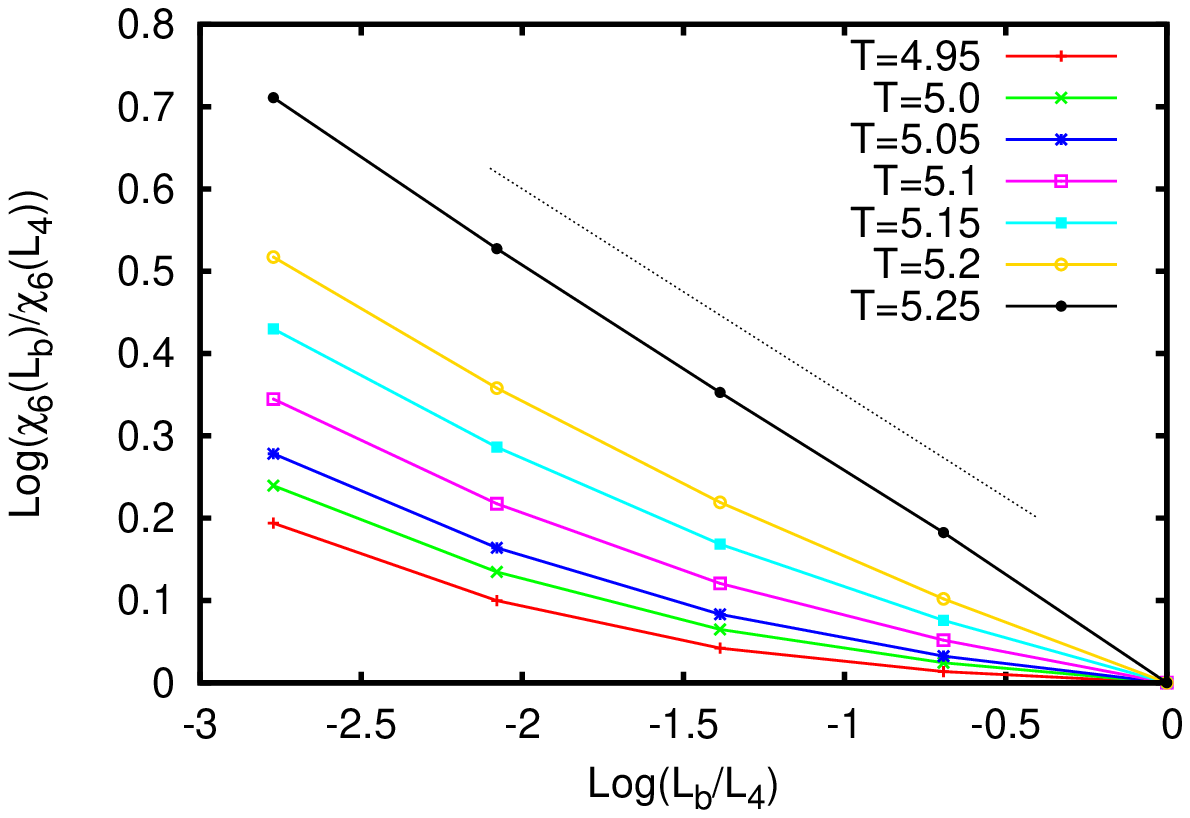}
\caption{Subblock scaling analysis for $\chi_6$, in one layer,  $\epsilon_w=5$. Dotted line corresponds to the slope $-1/4$,
the maximum possible slope for a hexatic phase. The slope of $\chi_6$ at $T=5.25$ is close to $-1/4$, for other
temperatures it is lower. The observed nonlinearity is due to defects.} 
\label{fig:scaling}
\end {figure}

Bagchi et al. \cite{bagchi96a} analyzed the scaling of the logarithm of the ratio $\chi_6(L_b)/\chi_6(L)$ versus $\ln(L_b/L)$. 
In the isotropic phase the slope should be $-2$ and in the hexatic phase 
it will be $-\eta_6\leq - 1/4$. For the crystal without
defects there should be no scaling. This relation is widely used to check the presence of the hexatic phase and finite size effects.
In Fig. \ref{fig:scaling} we present the results of the scaling for our system.
Due to defects in the crystal the scaling for low temperatures is not linear. The dotted line with the slope $-1/4$ 
reproduces the maximum slope expected in the hexatic phase.  As we can see from Fig.\ref{fig:scaling}, the slope of the
scaling curve for $T=5.25$ is very close to $1/4$, but we already know that
the corresponding $G_6$ decays on long distances exponentially, so, most
probably, at this temperature no hexatic phase exits. The scaling for lower temperatures does not allow us to distinguish
between a crystal with many defects and a probable hexatic phase. The scaling for $T=5.2$, where we expected the hexatic behavior, 
does not look any different from the one for $T=5.15$, that we verified as a crystal.

\begin{figure}[t]
\begin{center}
\includegraphics[width=8 cm]{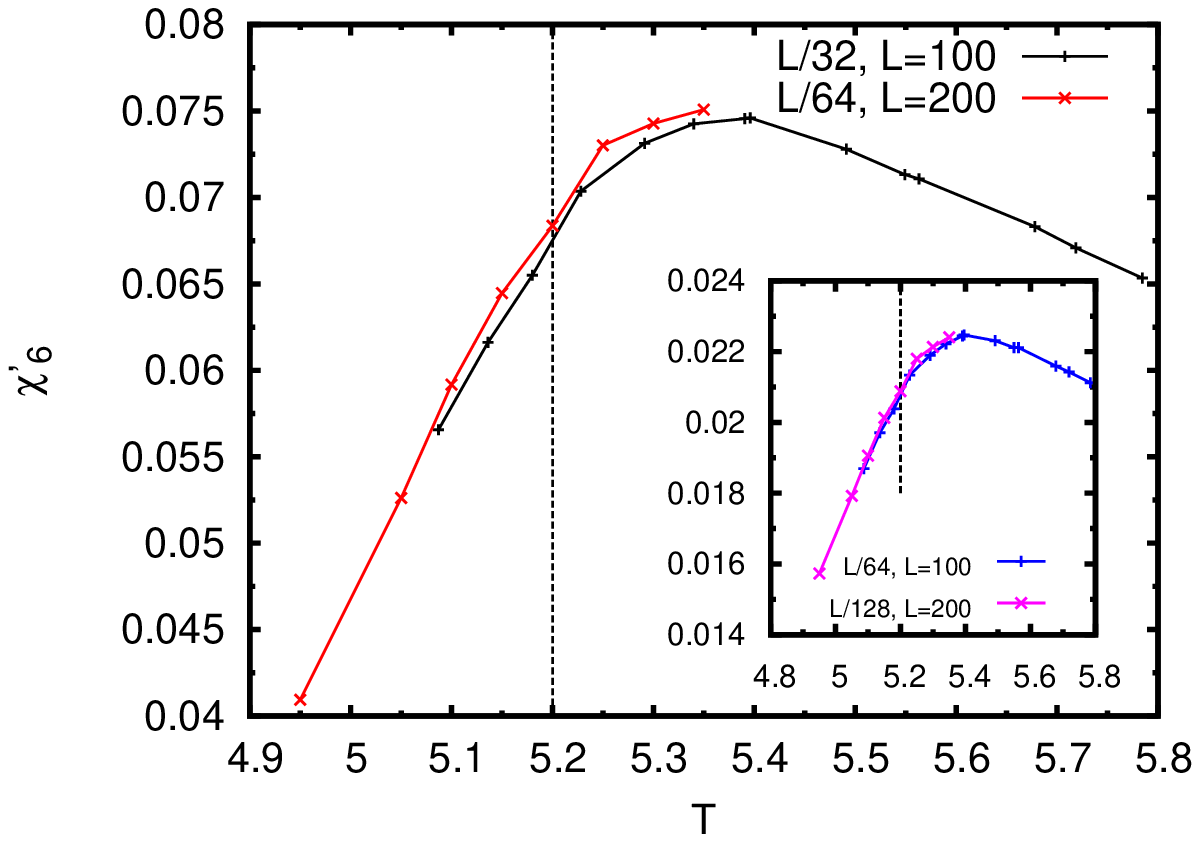}
\includegraphics[width=8 cm]{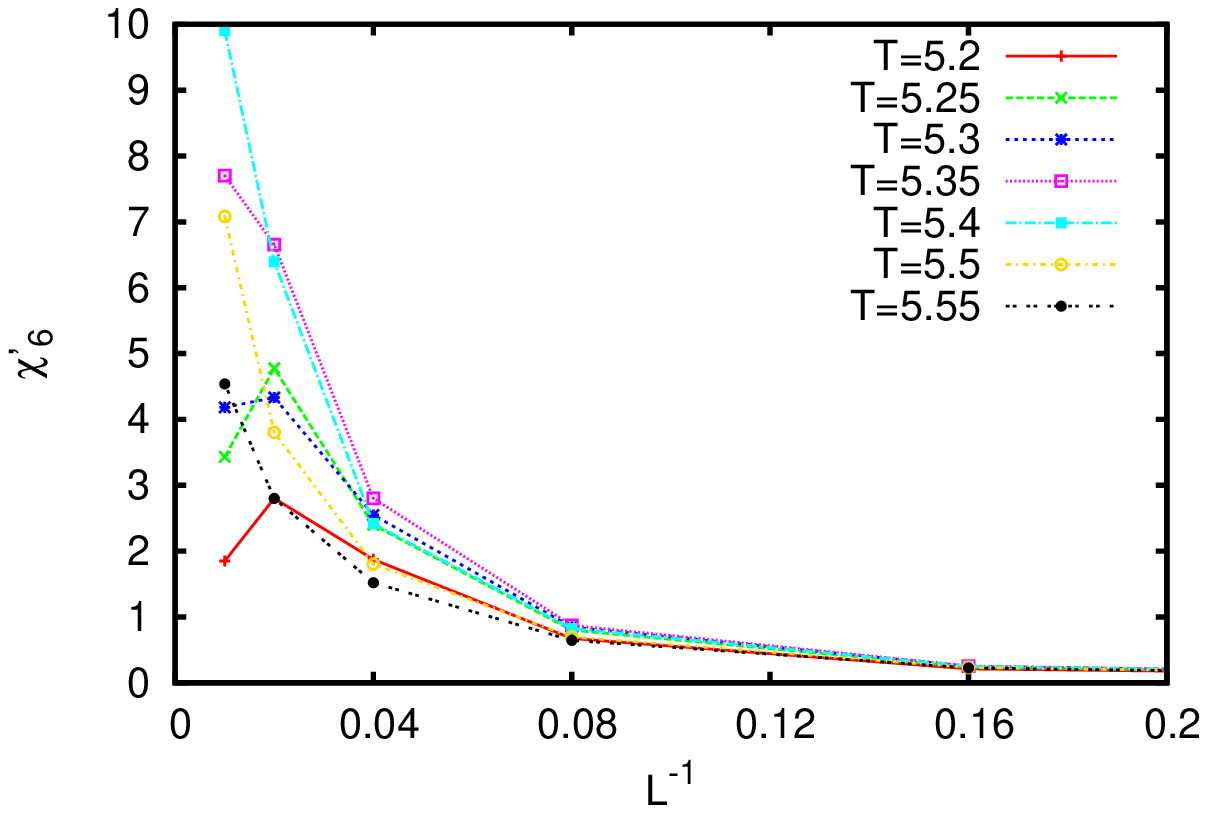}
\caption{Left figure: Fluctuation of the bond-order parameter, $\chi'_6$, as a function of temperature,
in one layer, $\epsilon_w=5$ for systems with box-lengths $100$ and $200$ for different subdivisions. 
The dashed line marks $T=5.2$, where we suspected a hexatic phase. The maxima of the curves are shifted to the
liquid region. Size effects are small.\\
Right figure: $\chi'_6$ as a function of the inverse value of the length of a subbox. 
The steepest slopes are observed for temperatures $5.35$ and $5.40$, both in the liquid phase.} 
\label{fig:chi_binder}
\end{center}
\end {figure}

Another version of susceptibility \cite{lee92a,weber95a}, $\chi'_6(L_B)$, measures the fluctuations of the bond-order parameter in 
the system:
\begin{equation}
k_B T\chi'_6=L^2 \left( \langle \Psi_6^2(L_b)\rangle -\langle \Psi_6(L_b) \rangle^2 \right),
\label{chi2}
\end{equation}
It should show a dramatic increase as the transition temperatures are approached either from solid or from liquid
phases and for the hexatic region it should become infinite \cite{weber95a}. However, it is impossible
to produce infinity in the simulations.
In Fig. \ref{fig:chi_binder} (left) the behavior of $\chi'_6$ as a function of temperature is presented.
We present our results both for our standard system with $L_y=L_z=200$ and a smaller one $L_y=L_z=100$ to
show that the size effects are very small. We compare $\chi'_6$ for
subdivisions $L/64$ and $L/128$ in the bigger system
that correspond to $L/32$ and $L/64$ in the smaller system. 
We can see that the maxima of all curves are shifted to the liquid phase to $T=5.35$, which is a 
consequence of the finite size effects in first order transitions \cite{weber95a,strepp01a}. The dashed
line marks the temperature $T=5.2$, where some signs of the hexatic phase were observed, here we do not
see any special features. The right part of Fig. \ref{fig:chi_binder} displays
the dependence of $\chi'_6$ on the inverse value of the length of a subbox.
If we do not subdivide the box, the scaling breaks down as we can clearly see on the graphs (first points).
This failure can be explained by the fact that we simulate in the canonical ensemble, but as soon as we start
subdividing the system, it behaves more like a grand-canonical ensemble. Unfortunately, we cannot extrapolate
our curves to $\chi_{\infty}$ as it was done in \cite{weber95a} since our way of subdivision does not provide 
enough data in the linear region of 
the curve. What we can still see is that the steepest slope is observed for temperatures $5.35$ and $5.40$, and
we already know both that temperatures lie in the liquid state region. We can
conclude that the observed transition is of a first order, but, as expected,
the transition temperatures obtained with this parameter are too high.

\begin{figure}[t]
\begin{center}
\includegraphics[width=8 cm]{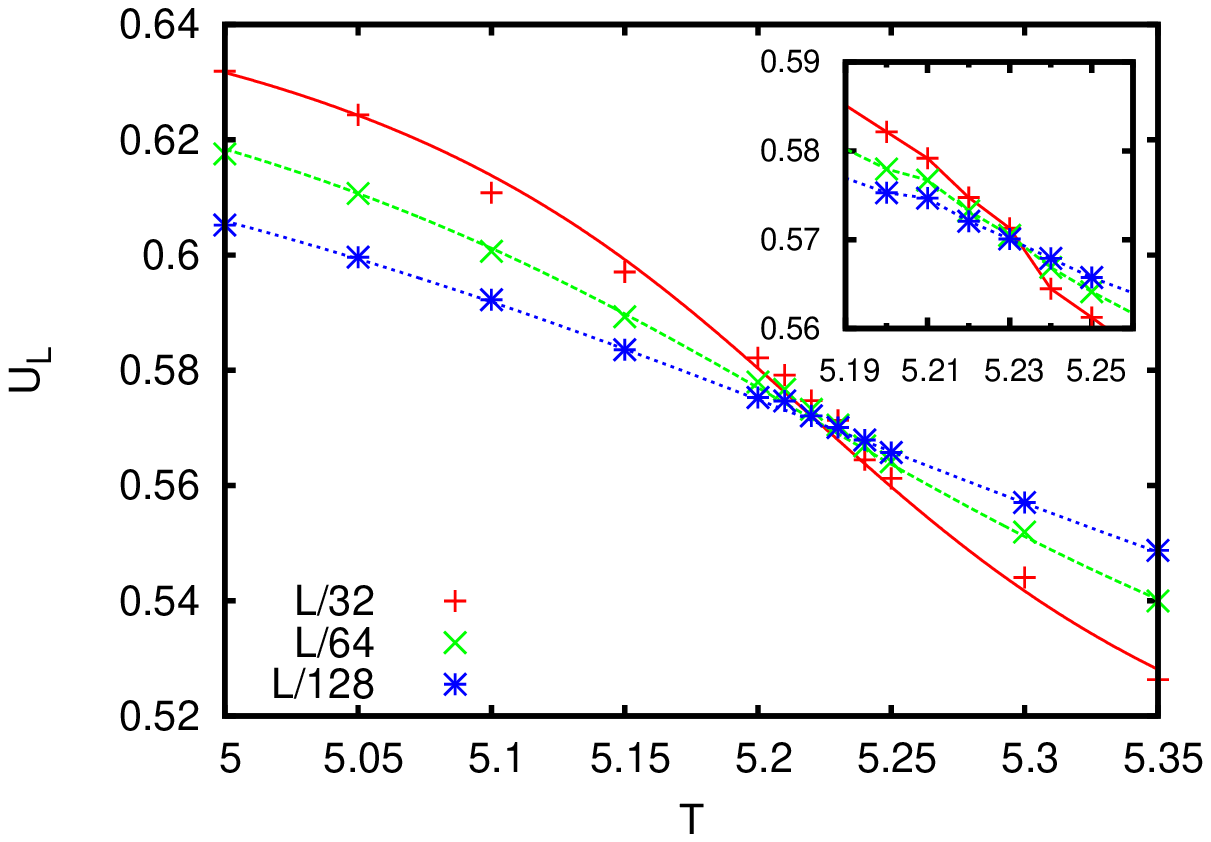}
\includegraphics[width=8 cm]{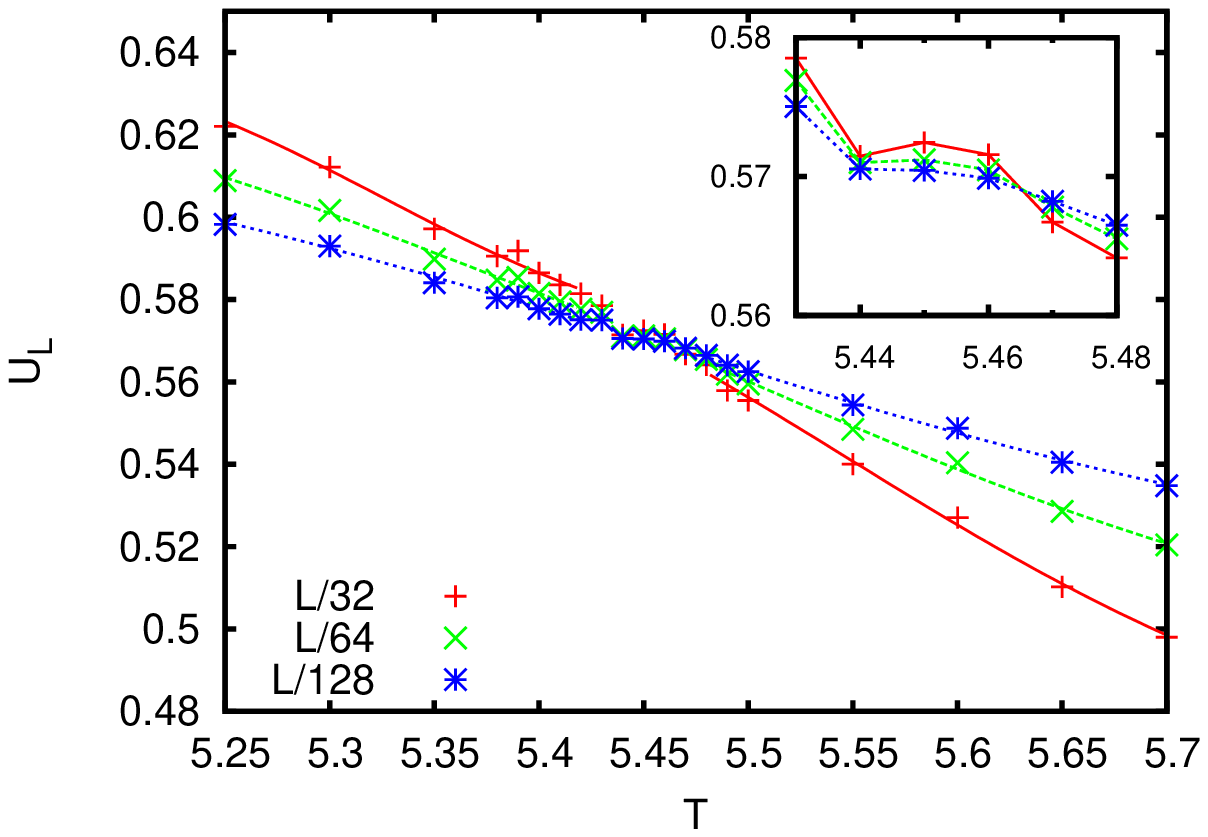}
\caption{Left figure: The Binder cumulant $U_l$ for several subdivisions of
  the system in one layer, $\epsilon_w=5$ as a function of temperature.
The curves do not collapse anywhere and intersect around $T=5.23$ (see inset), meaning that there is one  first 
order solid-liquid transition and no hexatic phase. Lines are guides to the eye.\\
Right figure: The Binder cumulant $U_l$ for several subdivisions of the system in one layer, $\epsilon_w=7$.
The curves are almost collapsing on one curve between temperatures $5.44$ and $5.46$ (see inset), a sign of
a possible hexatic phase. Lines are guides to the eye.} 
\label{fig:u_l}
\end{center}
\end {figure}

The last order parameter we investigate is the Binder cumulant \cite{binder81a,weber95a}:
\begin{equation}
U_L=1-\frac{\langle \Psi^4\rangle_6}{3\langle \Psi^2\rangle^2_6}.
\label{cumulant}
\end{equation}
Away from criticality in the limit of infinite system size, the cumulant  assumes different limiting values 
for ordered and disordered phases. For finite systems the value of the cumulant depends on the system size:
the smaller the system the more the cumulant deviates from the limiting value. In the case of a first order 
transition the cumulant exhibits an effective common intersection point at the
transition for sufficiently large systems. 
As for the hexatic phase the cumulant is expected to be independent of the system
size and to collapse onto one line over the entire range of the phase.
Fig. \ref{fig:u_l} (left) presents the behavior of the cumulant with temperature (lines serve as guides to the eye). 
We see clearly that there is
only one intersection point, meaning that there is one  first order solid-liquid transition and no hexatic phase.

Summarizing all our investigations we have shown that one should be quite
careful in choosing the order parameters in order to claim to have
observed a hexatic phase. In our case, the Binder cumulant provided the most stringent test.
The correlation function of the
bond-order parameter should be studied on big scales, since the crossover from the algebraic
decay to exponential can happen at relatively large distances. If scaling of the susceptibility 
does not give a straight line, that most probably means that we observe a very defective
crystal, but not a hexatic phase.

What would happen if the attraction of the walls becomes even more attractive? Let us look at the case $\epsilon_w=7$.
We will start from the last parameter we considered previously, $U_l$ given by
Eqn.(\ref{cumulant}), since we claimed that this was the most
sensitive parameter. As one can see in Fig.\ref{fig:u_l} (right), there is a small interval for temperatures between 5.44 
and 5.46 where the curves for different sub-divisions of the system  almost fall together. We interprete this as a sign of a
possible hexatic phase, since here the cumulant is independent of the system size. However, the temperature interval
is extremely narrow, so that the collapse of $U_l$ might not show a new phase
itself, but might be a precursor of a hexatic phase that would appear at higher
wall attraction or might happen only for infinite wall attraction.

\begin{figure}[t]
\includegraphics[width=8 cm]{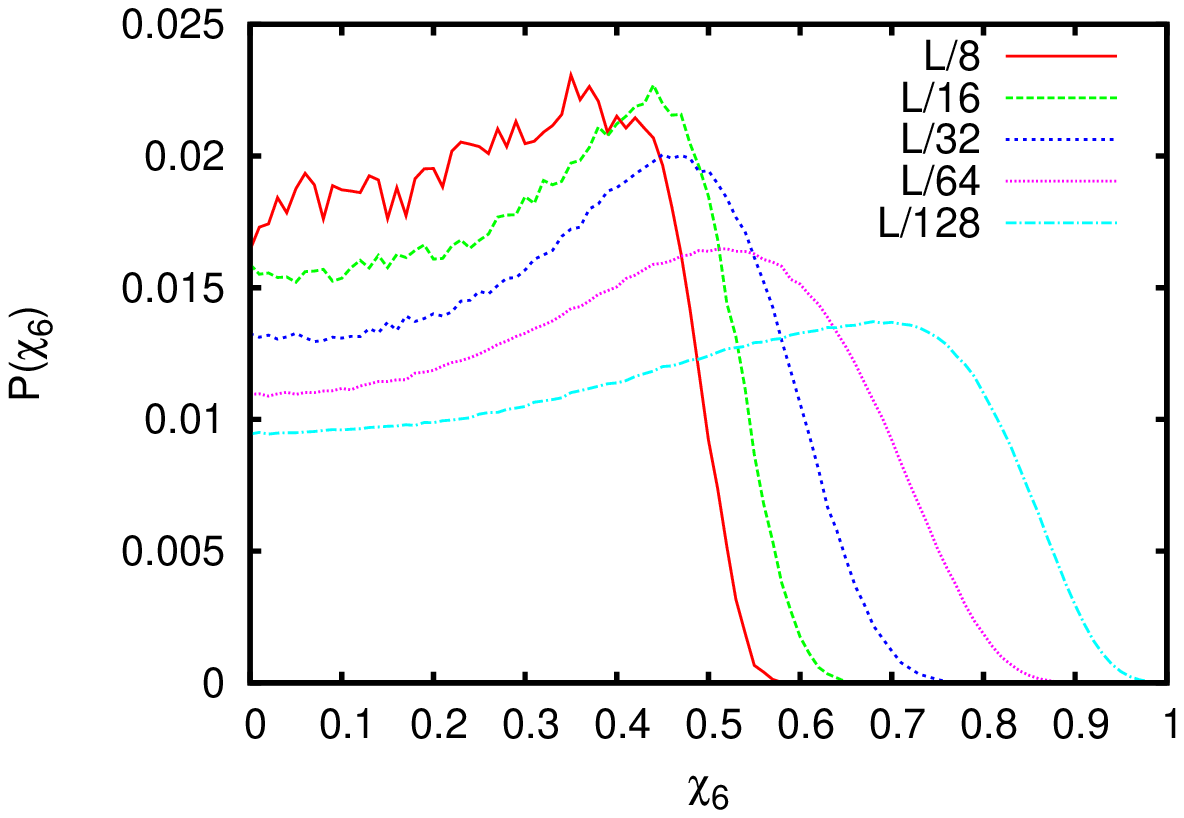}
\includegraphics[width=8 cm]{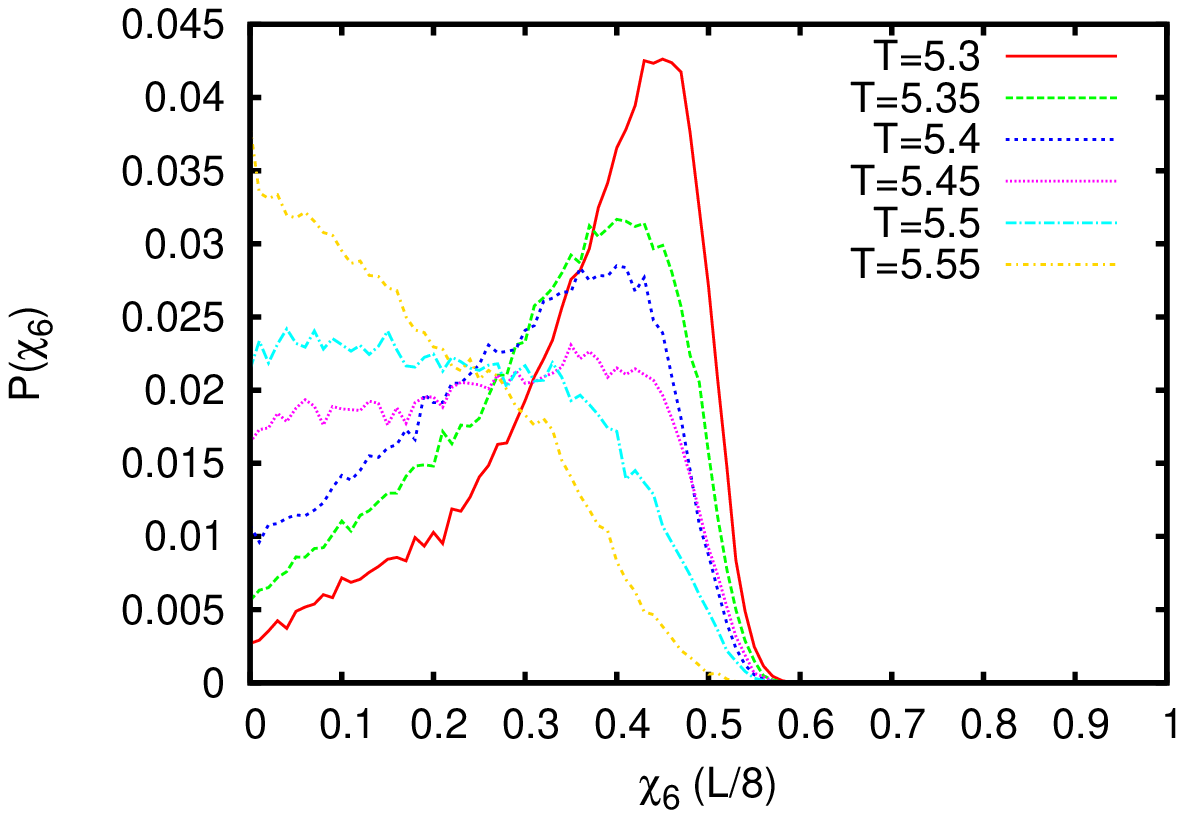}
\caption{Left figure: The distribution of $\chi_6$ for different subdivisions of the box in hexatic phase at $T=5.45$ 
in one layer, $\epsilon_w=7$. The distribution has one peak and the phase is homogeneous. \\
Right figure: The distribution of $\chi_6$ for subdivision $1/8$ at different temperatures. The peak of the
distribution gradually shifts away from $0.5$ and
becomes wider with increasing temperature and then at $T=5.55$ the system changes into the liquid state} 
\label{fig:distr1l7}
\end {figure}

The distribution of $\chi_6$ (\ref{chi1}) in the possible hexatic phase, $T=5.45$,  shows that the system is homogeneous 
(Fig. \ref{fig:distr1l7} left). If we look at the behavior of the $\chi_6$
distribution (Fig. \ref{chi1} right), taking as an example the
subdivison into $8\times8$ subblocks, we observe again that the peak of the
distribution decreases and slowly moves to lower values of $\chi_6$ with increasing temperature. 
We do not see any peculiarities in the distribution for $T=5.45$, and
at $T=5.55$ we observe a change to the characteristic liquid distribution with the maximum at $0$.

Also the behavior of other parameters, like the radial distribution function and 
the correlation of the bond-order parameter or scaling of $\chi_6$, 
does not  qualitatively differ from the behavior in the case of a less attractive wall $\epsilon_w=5$, so we go
through them briefly.

\begin{figure}[t]
\includegraphics[width=8 cm]{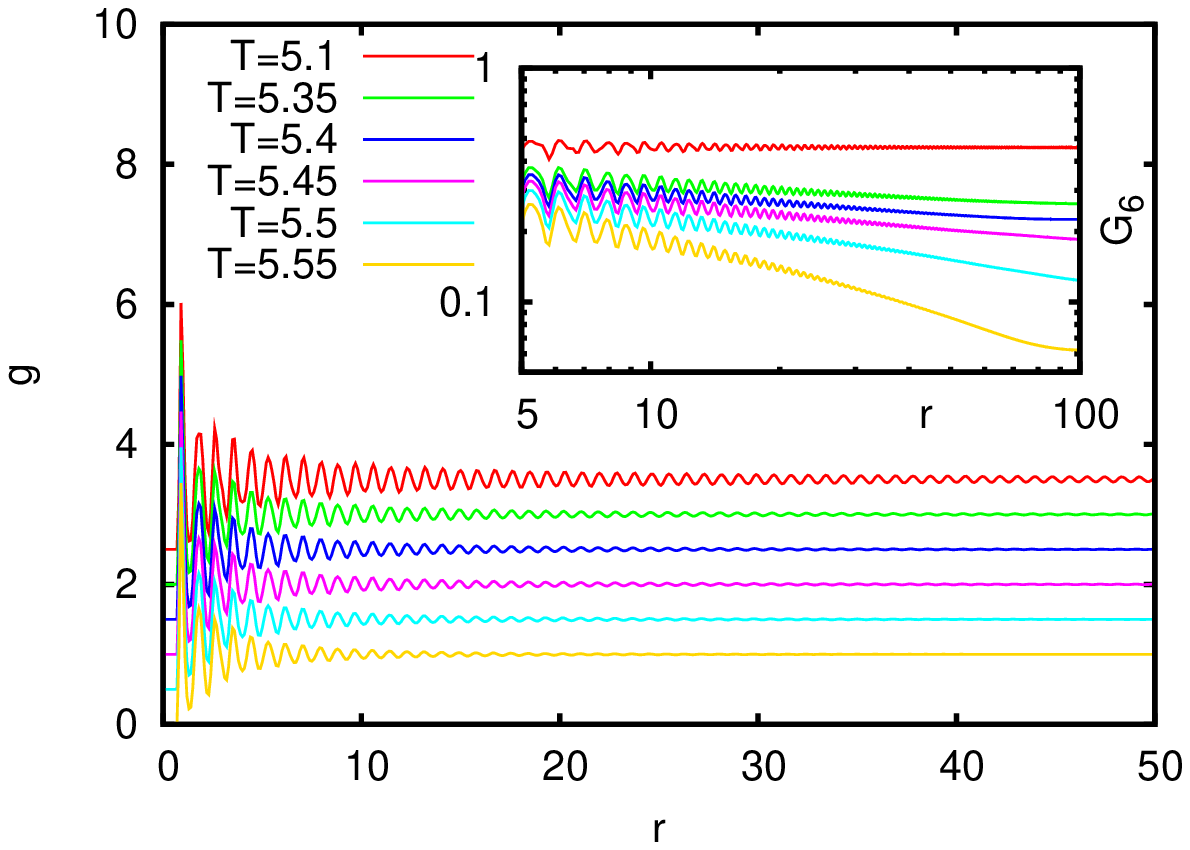}
\caption{Radial distribution function $g(r)$ for one layer, $\epsilon_w=7$. The curves are
shifted along the y-axis to separate them. RDF for $T=5.1$  is quasi long-ranged 
and solid-like, already at  $T=5.35$ it is short-ranged as for liquid.\\
Inset: Correlation of the bond-order parameter $g_6$ in one layer, $\epsilon_w=7$. For $T=5.1$ it does not decay,
for $T=5.35, 5.4, 5.45, 5.5$ it decays algebraically and for $T=5.55$ $G_6$ decays exponentially.} 
\label{fig:bond7}
\end {figure}

The correlation of the bond-order parameter decays algebraically for temperatures up to $T=5.5$ (inset of Fig. \ref{fig:bond7}),
and at $T=5.35$ and further on we see a short-range behavior of the radial distribution function (Fig. \ref{fig:bond7}).
However, we already know that for $T=5.35$ and $5.4$  there is no hexatic phase and 
we deal with a defective crystal. So, we conclude, that a combination of RDF
and $G_6$ is unreliable for claiming the occurence of a hexatic phase, since it does not 
distinguish it from a defective crystal.

\begin{figure}[t]
\includegraphics[width=8 cm]{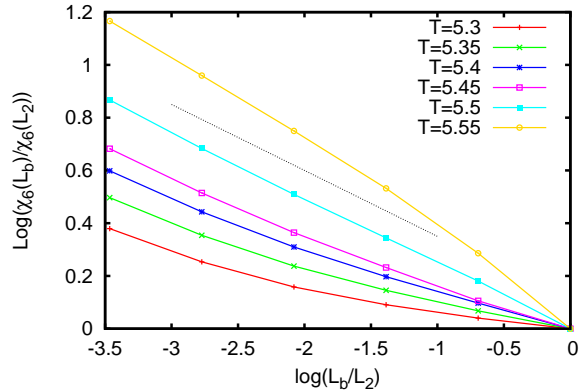}
\caption{Subblock scaling analysis for $\chi_6$, in one layer, $\epsilon_w=7$.  Dotted line corresponds to the slope $-1/4$,
the maximum possible slope for a hexatic phase.} 
\label{fig:chi7}
\end {figure}

The scaling of the $\chi_6(L_b)/\chi_6(L_0)$ for temperatures up to $T=5.45$ is below the $1/4$ 
slope (Fig. \ref{fig:chi7}). For $T=5.5$ the scaling of the ratio is exactly 0.25. The distribution of 
$\chi_6$ for $5.5$ is also not yet liquid (Fig. \ref{fig:distr1l7}). Therefore, according to distribution 
and scaling of $\chi_6$ and $G_6$  at the temperature $5.5$ the system has hexatic properties, but for the cumulant $U_l$ it
is already in the liquid phase. On the one hand, since we do not observe any two-phase 
region between a hexatic and liquid phase, one can assume that the Binder cumulant is oversensitive and can
omit some points. On the other hand, the transition from the hexatic to the liquid state proceeds very smoothly and 
the temperature $5.5$ can be treated as a boundary temperature between two phases. 

\begin{figure}[t]
\begin{center}
\includegraphics[width=8 cm]{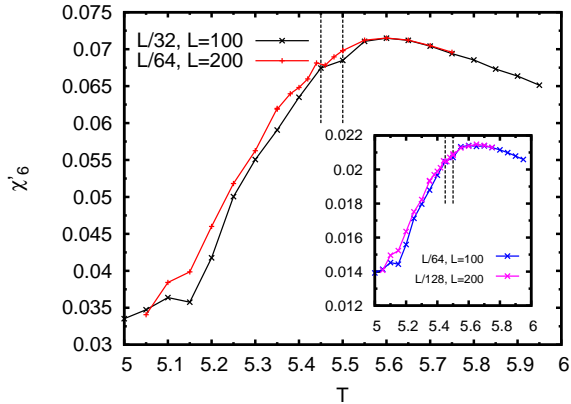}
\caption{Fluctuation of the bond-order parameter, $\chi'_6$, as a function of temperature,
in one layer, $\epsilon_w=7$, for systems with box-lengths $100$ and $200$ for
$L/32$ and $L/64$  subdivisions correspondingly. 
Dashed lines mark the region of a hexatic phase. Maximums of curves are shifted to the
liquid region. The size effects are small. Inset shows $\chi'_6$ for $L/64$
and $L/128$.} 
\label{fig:chi_binder7}
\end{center}
\end {figure}

If we look at the change of $\chi'_6(L)$ (\ref{chi2}) with temperature, we observe maxima between the temperatures
$5.6$ and $5.65$ (Fig. \ref{fig:chi_binder7}), which indicate that the transition to the liquid is of first order,
since the maxima for this parameter are far in the liquid phase. However, no change in behavior is observed
in the possible region of a hexatic phase, the borders  of which are marked
with dashed lines. We compare again the behavior of $\chi'_6(L)$
for subdivisions $L/64, L/128$ in our standard system with
box-length $200$ and  $L/32, L/64$ in a smaller one with $L_y=L_z=100$. The curves for the
corresponding subdivisions coincide in the solid without defects and the liquid region. They show a small quantitative
difference  when the crystal gains defects and then goes to the hexatic phase, which region is marked by two dashed lines.

We have shown that only the Binder cumulant
  behaves qualitatively different for the systems with $\epsilon_w=5$
  and $\epsilon_w=7$. The behavior of the other parameters is similar
  for both cases. In our simulations, the Binder cumulant shows signs
  of a hexatic phase only at the strongest wall attraction
  $\epsilon_w=7$. We checked the fluctuations of particles
  perpendicular to the walls by fitting the density profile
  to a Gaussian distribution. The fluctuation for $\epsilon_w=5$ is
  $\Delta z\approx 0.175$ and $\Delta z\approx 0.15$ for
  $\epsilon_w=7$, practically independent of the temperature in the
  studied range. Both values are significantly smaller than $0.4$,
  which was reported to be the maximally possible fluctuations for
  liquid-to-hexatic transition in the quasi-2D system of hard
  spheres\cite{xu2008}. Since our Lennard-Jones particles are much
  softer than hard spheres and therefore can overlap to some extent,
  this might explain why we need a stricter confinement to observe a hexatic phase.

\section{Conclusions} \label{sec:conclusions}

We carried out a simulation study of the liquid-to-solid
transformation of a LJ fluid in a slit pore accommodating one or two layers
and several ratios of wall-particle to particle-particle attractions.
To investigate the possible existence of an intermediate hexatic phase that was
previously observed not only in 2d, but also in quasi-2d systems \cite{binder02a,Radhakrishnan2002,xu2008}, 
we performed an analysis using a broad range of order parameters. 
Studying the radial distribution function together with the decay of the
bond-order correlation turned out to be insufficient, since, on the one hand, 
they characterize the system in a whole, so one can not conclude from their behavior if
the system is homogeneous.
On the other hand, they are also insensitive to defects, so one 
cannot distinguish properly between a defective crystal and a hexatic phase.
The scaling of different parameters turned out to be much more successful.
With the help of the angular susceptibility one can easily check the
homogeneity of the system. We consider the Binder cumulant to be the most reliable
parameter in distinguishing a hexatic phase, since it shows either existence
or absence of it. In our study 
the temperature of transition computed with the help of $U_L$ does not coincide 
with the temperature where the fluctuation of the bond-order 
parameter has a maximum, which makes it possible to assume a first order transition
in all cases \cite{weber95a}. We observed  signs of a possible intermediate hexatic
phase only in the slit with extremely attractive walls and a single layer
of particles, i. e. if the system is practically 2d, otherwise there is
a single liquid-solid transition.
This is in contrast to the works \cite{Radhakrishnan2002,radhakrishnan2002JCP}
  on a similar system, in which signs of a hexatic phase in the contact layers near the wall
  were observed even in systems with up to seven layers, at wall strengths comparable to our $\epsilon_w=7$.
  These findings, however, were based on studying the behavior of global order
  parameters and scaling of $\Psi_6$ only,
  which, as we have shown, are not sufficient to safely detect a hexatic phase.

Our results have implications for experimental studies on the hexatic phase~\cite{murray87,zahn1998},
  since they show that even a monolayer requires a strong confining
  force to exhibit a true hexatic phase, while studies based on the decay of the RDF and $G_6$ can easily be fooled by
  defective crystals or coexistent phases. This might explain why some studies find a hexatic phase,
  while other studies of a seemingly very similar system do not.

\begin{acknowledgments}
The authors thank K. Binder for valuable remarks. NG benefited from  fruitful
discussions with  E.E. Tareyeva, V.N. Ryzhov and M. Sega. 
We thank the DFG for financial support through the SPP 1296.
\end{acknowledgments}

\bibliographystyle{aipsamp}

\end{document}